\documentclass[aip,jmp,12pt,showpacs,showkeys]{revtex4-1}

\usepackage{graphicx}
\usepackage{epsfig}
\usepackage{amssymb}
\usepackage{amsbsy}
\usepackage{amsmath}
\usepackage{bm}

\newcommand{\Def}{\stackrel{\textrm{\footnotesize def}}{=}}
\def\N{\mathsf{N}}
\def\G{\mathsf{G}}
\def\U{\mathcal{U}}
\def\V{\mathcal{V}}
\def\X{\mathcal{X}}

\def\s{\widehat{\sigma}}
\def\x{\widehat{x}}

\begin{document}
\title{Conjugate Flow Action Functionals}

\author{Daniele Venturi}
\email{daniele\_venturi@brown.edu}
\noaffiliation
\affiliation{Division of Applied Mathematics, Brown University}
\date{\today}

\begin{abstract}
We present a new general method to construct 
an action functional for a non-potential field 
theory. The key idea relies on representing the 
governing equations of the theory relative 
to a diffeomorphic flow of curvilinear coordinates 
which is assumed to be functionally dependent 
on the solution field.
Such flow, which will be called the 
conjugate flow of the theory, evolves in space 
and time similarly to a physical fluid 
flow of classical mechanics and it can be selected in 
order to symmetrize the G\^ateaux derivative of the 
field equations with respect to suitable local bilinear forms. 
This is equivalent to requiring that the governing equations 
of the field theory can be derived from a 
principle of stationary action on a Lie group manifold. 
By using a general operator framework, we obtain the 
determining equations of such manifold and the 
corresponding conjugate flow action functional.
In particular, we study scalar and vector 
field theories governed by second-order nonlinear 
partial differential equations. 
The identification of transformation groups
leaving the conjugate flow action functional invariant 
could lead to the discovery of new conservation laws in 
fluid dynamics and other disciplines.
\end{abstract}




\maketitle

\section{Introduction}
\label{sec:intro}

The problem of finding an action functional whose 
Euler-Lagrange equations correspond to a prescribed 
set of partial differential equations (PDEs) is known as the 
inverse problem of the calculus of variations and it has attracted the 
attention of researches for more than one century \cite{Filippov,Berdi}.
Perhaps, one the main reasons is that the formulation of 
a field theory in terms of an action functional is very 
elegant and, more importantly, it allows us to establish 
an immediate connection between symmetry principles and 
conservation laws \cite{Lovelock,Noether,Landau,Weinberg}.

For a given system of field equations, the existence conditions 
of an action functional can be put in a correspondence with the theory of 
conservative vector fields \cite{Magri,Tonti1,Tonti4}. 
Essentially, if the path integral of the nonlinear operator 
representing the field equations is independent on the 
trajectory of fields connecting two specified points in a function 
space, then there exists a scalar field (the action) whose 
functional derivative yields the governing equations of the theory. 
The path integral of an operator along a trajectory of fields 
is defined in terms of a bilinear form \cite{Magri_initial,Tonti3}
which can be selected in a rather arbitrary way. 
In this sense, the solution to the inverse problem of the 
calculus of variations is reduced to look for a bilinear 
form that makes the given nonlinear operator potential\cite{Filippov}, 
if any. 
It was shown by Tonti \cite{Tonti} that there exist not just 
one but an infinite number of such forms. 
Therefore, an infinite number of action principles 
can be constructed for a given set of field equations. 
However, the physical meaning of such generalized principles 
is often obscured by the bilinear form that 
has to be adjusted on the given system of field equations. 
An alternative way to proceed is to select a specific bilinear 
form, e.g., one having a physical meaning, and then look for ways of 
modifying the field equations as to obtain a new problem which 
is potential with respect to the chosen bilinear form. 
Among known methods devised to do so, 
we recall the adjoint equation method \cite{Morse,Finlayson,Martin,Jouvet} 
and the integrating operator method \cite{Tonti}.

The purpose of this paper is to introduce a new general approach 
to construct an action functional for a non-potential field theory. 
The key idea relies on representing the
governing equations of the theory relative to a curvilinear flow 
of coordinates that is assumed to be a functional of the solution field. 
This flow will be called the {\em conjugate flow} of the theory and, 
as we will see, it can be selected in order to satisfy the existence 
conditions of an action principle for any given set of field equations. 
Let us briefly describe the main ideas that led us to introduce 
the conjugate flow and, more importantly, their relevance 
in the context of known physical theories. 
To this end, let us first notice that flows of coordinates depending 
on solution to a system of field equations arise naturally 
in many areas of mathematical physics. 
Perhaps, the most relevant example is in the context of 
classical fluid mechanics, where the trajectories of 
fluid elements in space are related to the velocity field that 
solves, e.g., the Navier-Stokes equations \cite{Aris,Batchelor,Spurk}. 
The curvilinear coordinate system advected by such flow 
is known as {\em Lagrangian system}
and it can be determined by integrating out the definition of the 
velocity field \cite{Venturi_A}. In this sense, physical 
fluid flow of classical mechanics can be considered 
as a very particular type of conjugate flow.
Another example of conjugate flow is the 
free-falling coordinate system \cite{Weinberg} 
in the Einstein's theory of gravitation. Here the flow 
appears as a geodesic mapping \cite{Lovelock} in a four-dimensional 
Riemannian space whose metric is determined by a particular 
distribution of energy and momentum through the solution of the 
Einstein's field equations. 

In both examples just discussed, the relation between the 
solution to the field equations and the conjugate flow 
is given and it reduces to the definition of the 
velocity field in the case of Navier-Stokes equations and 
to the definition of geodesics in the 
case of Einstein's theory of gravitation. 
In a broader framework, however, 
such functional relation may be left unspecified. 
This yields an infinite number of functional degrees of 
freedom (those associated with the conjugate flow) that
can be selected, for instance, by requiring that the 
governing equations of the theory can be derived 
from a principle of stationary action. 
This new formulation of the inverse problem of the calculus 
of variations brings together concepts of differential 
geometry and nonlinear functional analysis and it 
results in {\em new types of action principles} 
generalizing those ones based on specific functional 
flows, e.g., the Herivel-Lin principle for perfect 
fluids \cite{Herivel,Eckart,Bretherton,Morrison}.

This paper is organized as follows. In section \ref{sec:conj} we introduce 
the theory of the conjugate flow and we characterize the group of 
infinitesimal perturbations by using methods of nonlinear functional analysis.
The representation of arbitrary nonlinear field equations in 
conjugate flow intrinsic coordinates is discussed 
in section \ref{sec:rep}. In section \ref{sec:action} we determine the 
existence conditions of conjugate flow action functionals in a rather general 
operator framework. Determining equations for conjugate flows 
symmetrizing scalar and vector field theories governed by second-order PDEs 
are obtained in section \ref{sec:formal symmetry}.
Finally, the main findings and their implications are 
summarized in section \ref{sec:summary}. We also include a brief 
appendix where we recall some fundamental identities of
differential geometry that will be extensively used throughout 
the paper.

\section{The conjugate flow}
\label{sec:conj}
\begin{figure}
\centerline{\includegraphics[height=7cm]{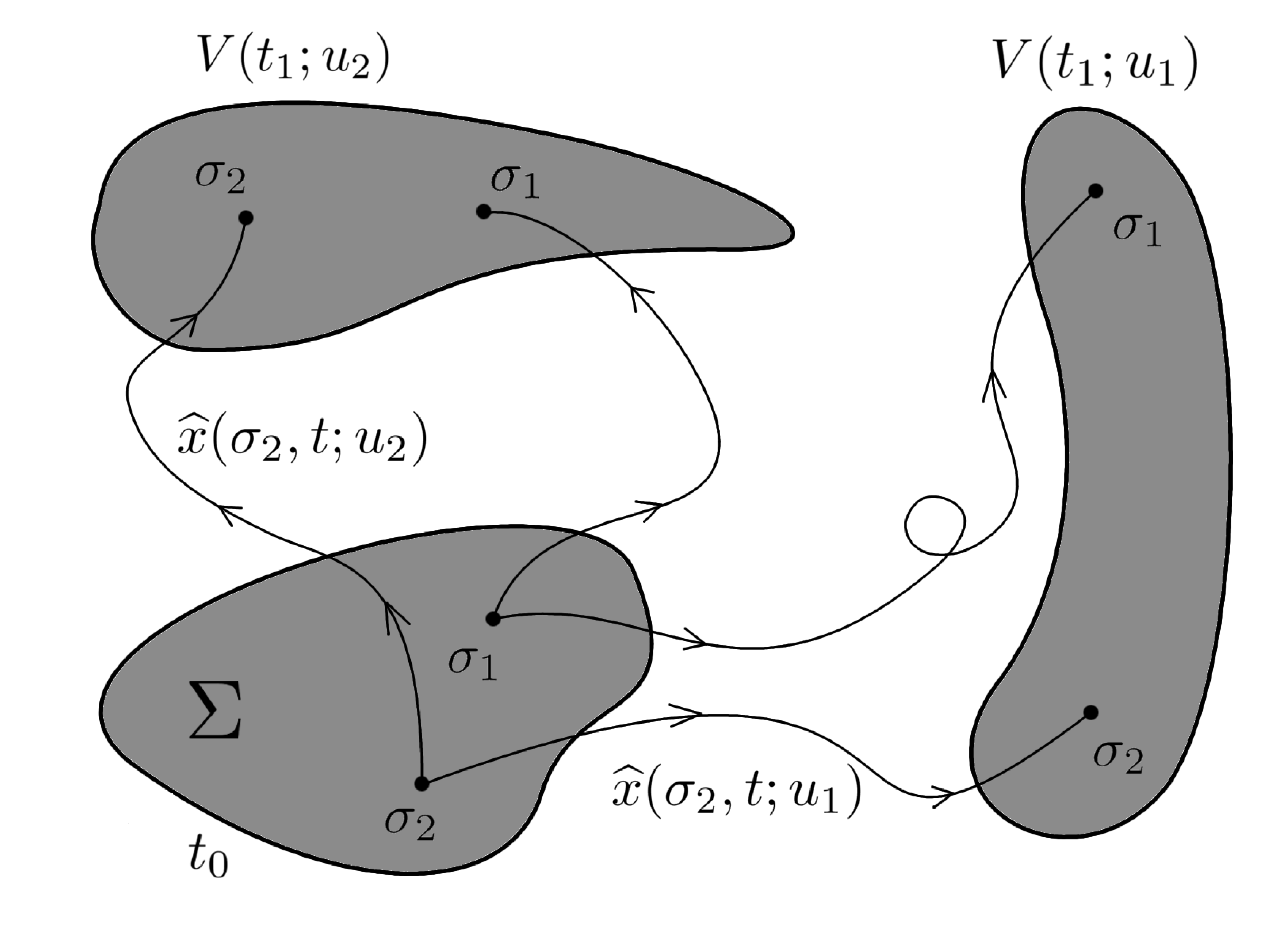}}
\caption{Sketch of the transport phenomenon of a volume 
of particles $\Sigma$ by two different realizations of the conjugate 
flow $\x^\mu(\sigma^\nu;u^j)$, namely, $\x^\mu(\sigma^\nu;u_1^j)$ 
and  $\x^\mu(\sigma^\nu;u_2^j)$.  Shown 
are trajectories of two different particles labeled
as $\sigma_1$ and $\sigma_2$.}
\label{fig:two_conj_flows}
\end{figure}

Conjugate flow is an intuitive physical notion which 
is represented mathematically by a continuous point transformation 
of Euclidean or Riemannian space into itself. In order to set up this
transformation, let us consider a ``particle'' labeled 
by $\sigma^\nu$ ($\nu=0, 1,...,n$, where $n$ is the number of spatial 
dimensions and $0$ denotes the temporal component) 
and represent its trajectory in a fixed 
space-time Cartesian system as
\begin{equation}
 x^\mu_\sigma=\x^\mu\left(\sigma^\nu;u^j\right),\quad\quad\mu,\nu=0,..,n,
 \label{tr_conj}
\end{equation}
where $u^j$ ($j=1,...,N$) is a vector field that solves a 
prescribed system of field equations.
We adopt the convention that 
Latin indices $i$, $j$, $k$, etc., run over spatial coordinate 
labels (usually $1$, $2$, $3$) while Greek indices $\mu$, $\nu$, $\alpha$, etc., 
run over space-time coordinate labels (usually $0$, $1$, $2$, $3$). 
Also, repeated indices are summed unless otherwise stated. 
The trajectory of $\sigma^\nu$ defined by (\ref{tr_conj}) can 
be of course expressed in arbitrary coordinate 
systems \cite{Venturi_A,Lovelock,Weinberg}, 
although such generalization is not of primary 
importance in what follows. If we consider three spatial 
dimensions and we assume that the time variable is not 
transformed, i.e.,  $\x^0=\sigma^0=t$, then the system (\ref{tr_conj}) 
reduces to the more familiar form \cite{Aris,Batchelor,Spurk}
\begin{eqnarray}
x_\sigma^1=\x^1\left(\sigma^1,\sigma^2,\sigma^3,t;u^j\right)\nonumber\\
x_\sigma^2=\x^2\left(\sigma^1,\sigma^2,\sigma^3,t;u^j\right)\nonumber\\
x_\sigma^3=\x^3\left(\sigma^1,\sigma^2,\sigma^3,t;u^j\right)\nonumber
\label{tr_conj_C}
\end{eqnarray}
where $\sigma^j=\x^j\left(\sigma^1,\sigma^2,\sigma^3,t_0;u^j\right)$ is
the initial position of the particle.
The transformation (\ref{tr_conj}) is assumed 
to be invertible (with differentiable inverse) and 
to possess continuous derivatives up to a prescribed 
order, except possibly at certain singular surfaces, curves or points. 
These requirements make (\ref{tr_conj}) a {\em diffeomorphism}, i.e., 
a time-dependent flow of curvilinear coordinates
\cite{Venturi_A,Matolcsi,Thiffeault,Thiffeault1,Haoxiang} 
whose motion in space resembles \emph{in toto} a 
physical fluid flow of classical mechanics. 
Einstein called these coordinate systems
``reference-mollusks'' \cite{Einstein}.
In figure \ref{fig:two_conj_flows} we sketch 
two realizations of the conjugate flow (\ref{tr_conj}) 
for two different solution fields corresponding, e.g., to 
different boundary or initial conditions in an initial/boundary
value problem for a field equation.

Coordinate flows that are functionals of the solution to a
field equation are obviously not new in the literature. 
For instance, in the context of symmetry 
analysis of partial differential equations the so-called 
{\em non-classical symmetries} \cite{Bluman} are remarkable examples of 
field-dependent transformations.
Similarly, in classical Lagrangian fluid dynamics 
the trajectories of fluid elements in space are obtained as 
{\em local functionals} of the velocity field $U^j\left(x^k,t\right)$ that 
solves, e.g., the Navier-Stokes equations ($x^k$ here are fixed Cartesian
coordinates)
\begin{eqnarray} 
 \frac{\partial U^j}{\partial  t}+U^k\frac{\partial U^j}{\partial x^k}=
 -\frac{\partial P}{\partial x^k}+\frac{1}{\textit{Re}}\frac{\partial^2 U^j}{\partial x^k\partial x^k},
 \quad\quad\frac{\partial U^k}{\partial x^k}=0,\quad\quad k,j=1,..,n.\label{NS}
\end{eqnarray}
Such functional relation is defined by the solution to the well-known 
initial value problem
\begin{equation}
 \frac{\partial \widehat{X}^j(\sigma^i,t)} {\partial t}=U^j\left(\widehat{X}^k(\sigma^i,t),t\right),
 \quad \quad \widehat{X}^j(\sigma^i,t_0)=\sigma^j.\label{frel}
\end{equation}
In this sense, the physical fluid flow can be considered as a 
very particular type of conjugate flow, since $\widehat{X}^j$ is functionally 
dependent on the solution to Eq. (\ref{NS}) by means of Eq. (\ref{frel}). 
Another conjugate flow which is different than the physical fluid 
flow may be defined, e.g., by solving
\begin{equation}
 \frac{\partial \widehat{X}^j_\omega(\sigma^i,t)} {\partial t}
 =U^j\left(\widehat{X}^k_\omega(\sigma^i,t)+B^k_\omega(\sigma^i,t),t\right),\quad \quad 
 \widehat{X}^j_\omega(\sigma^i,t_0)=\sigma^j.
 \label{frel3}
\end{equation}
where $B^k_\omega(\sigma^i,t)$ is a realization of 
space-time Brownian motion. A remarkable result 
by Gomes \cite{Gomes} shows that ensemble averaging 
- over random $B^k_\omega$ - of diffeomorphisms of type (\ref{frel3}) 
allows to construct a variational principle for the Navier-Stokes 
equations.

\subsection{Infinitesimal flow perturbations}
The components of the vector field $u^j$ appearing 
in Eq. (\ref{tr_conj}) are, by definition, Cartesian 
components expressed in terms of conjugate flow intrinsic 
coordinates $\sigma^\nu$. In other words, if we denote 
by $U^j(x^\mu)$ the Cartesian components of a vector 
field that satisfies, e.g.. Eq. (\ref{NS}), then $u^j$ are defined as
\begin{equation}
 u^j(\sigma^\nu) \Def U^j(\x^\mu(\sigma^\nu;u^k)).
 \label{field}
\end{equation}
Note that these are {\em not} the tensorial components 
\cite{Aris,Truesdell1} of the vector field.
Now, let us consider an infinitesimal perturbation of $u^j$ in the form 
\begin{equation}
  u^j(\sigma^\nu) \rightarrow u^j(\sigma^\nu)+\epsilon\varphi^j(\sigma^\nu),
 \label{field_perturb}
\end{equation}
where $\epsilon$ is a small real parameter. 
Such perturbation can be obviously expressed relative 
to arbitrary coordinate systems. For instance, in Cartesian 
coordinates we have $U^j(x^\mu)\rightarrow U^j(x^\mu)+\epsilon\Phi^j(x^\mu)$. 
Disregarding the particular choice of the coordinate system, 
the field perturbation (\ref{field_perturb}) induces the following perturbation
in the conjugate flow (\ref{tr_conj}) 
\begin{equation}
\x^\mu(\sigma^\nu;u^j+\epsilon \varphi^j)\simeq\x^\mu(\sigma^\nu;u^j)+
\epsilon\frac{\delta \x^\mu}{\delta u^j}\varphi^j,
\label{xper}
\end{equation}
where, by definition
\begin{equation}
 \frac{\delta \x^\mu}{\delta u^j}\varphi^j\Def
 \lim_{\epsilon\rightarrow0}\frac{\x^\mu(\sigma^\nu;u^j+\epsilon \varphi^j)-
 \x^\mu(\sigma^\nu;u^j)}{\epsilon}.
 \label{G1}
\end{equation}
The quantity $\delta\x^\mu/\delta u^j$ is known as G\^ateaux 
derivative \cite{Vainberg} of the functional $\x^\mu$ with respect 
to $u^j$ and, under rather weak requirements \cite{Nashed}, it is a 
continuous linear operator.
The perturbed flow $\widehat{x}^\mu(\sigma^\nu;u^j+\epsilon\varphi^j)$ 
is assumed to have the same regularity properties 
as the unperturbed one, i.e., invertibility and continuous 
derivatives up to prescribed order in all variables. 

Let us now {\em postulate} that the solution field $u^j$ is also
functionally connected to the conjugate flow $\x^\mu$ and let 
us denote this functional relation by $u^j(\sigma^\nu;\x^\mu)$. 
This fundamental assumption implies that an 
infinitesimal flow perturbation $\x^\mu+\epsilon\widetilde{\phi}^\mu$ 
induces the following variation in the solution field $u^j$
\begin{equation}
u^j(\sigma^\nu;\x^\mu+\epsilon \widetilde{\phi}^\mu)\simeq u^j(\sigma^\nu;\x^j)+
\epsilon\frac{\delta u^j} {\delta\x^\mu}\widetilde{\phi}^\mu,
\label{uper}
\end{equation} 
where, in analogy with Eq. (\ref{G1}), we have defined the 
G\^ateaux differential as
\begin{equation}
\frac{\delta u^j} {\delta\x^\mu}\widetilde{\phi}^\mu\Def\lim_{\epsilon\rightarrow0}
\frac{u^j(\sigma^\nu;\x^j+\epsilon \widetilde{\phi}^\mu)-u^j(\sigma^\nu;\x^j)} {\epsilon}.
\end{equation}
In the context of the Navier-Stokes equations, 
this means that a perturbation in the conjugate flow $\x^j$ determines 
- by assumption - a perturbation in the velocity field $U^j$ 
that solves Eq. (\ref{NS}). This ultimately results in a perturbation 
of the physical fluid flow $\widehat{X}^j$ by means of Eq. (\ref{frel}). 
In other words, by perturbing the conjugate flow we are actually 
perturbing the physical fluid flow.
At this point it is convenient to set
\begin{eqnarray}
\widetilde{\varphi}^j&=&\frac{\delta u^j} {\delta \x^\mu}\widetilde{\phi}^\mu,\label{conj2}\\
\widehat{\phi}^\mu&=&\frac{\delta \x^\mu}{\delta u^j}\varphi^j\label{conj1}
\end{eqnarray}
and write Eq. (\ref{xper}) and Eq. (\ref{uper}) as
\begin{eqnarray}
u^j(\sigma^\mu;\widehat{x}^\nu+\epsilon\widetilde{\phi}^\nu)&\simeq 
u^j\left(\sigma^\mu;\widehat{x}^\nu\right)+
\epsilon\widetilde{\varphi}^j\left(\sigma^\mu;\widehat{x}^\nu\right),\label{field_perturbation}\\
\widehat{x}^\mu\left(\sigma^\nu;u^j+\epsilon \varphi^j\right)&\simeq\x^\mu\left(\sigma^\nu;u^j\right)+
\epsilon\widehat{\phi}^\mu\left(\sigma^\nu;u^j\right).\label{trajectory_perturbation}
\end{eqnarray}
\begin{figure}
 \centerline{\includegraphics[height=7cm]{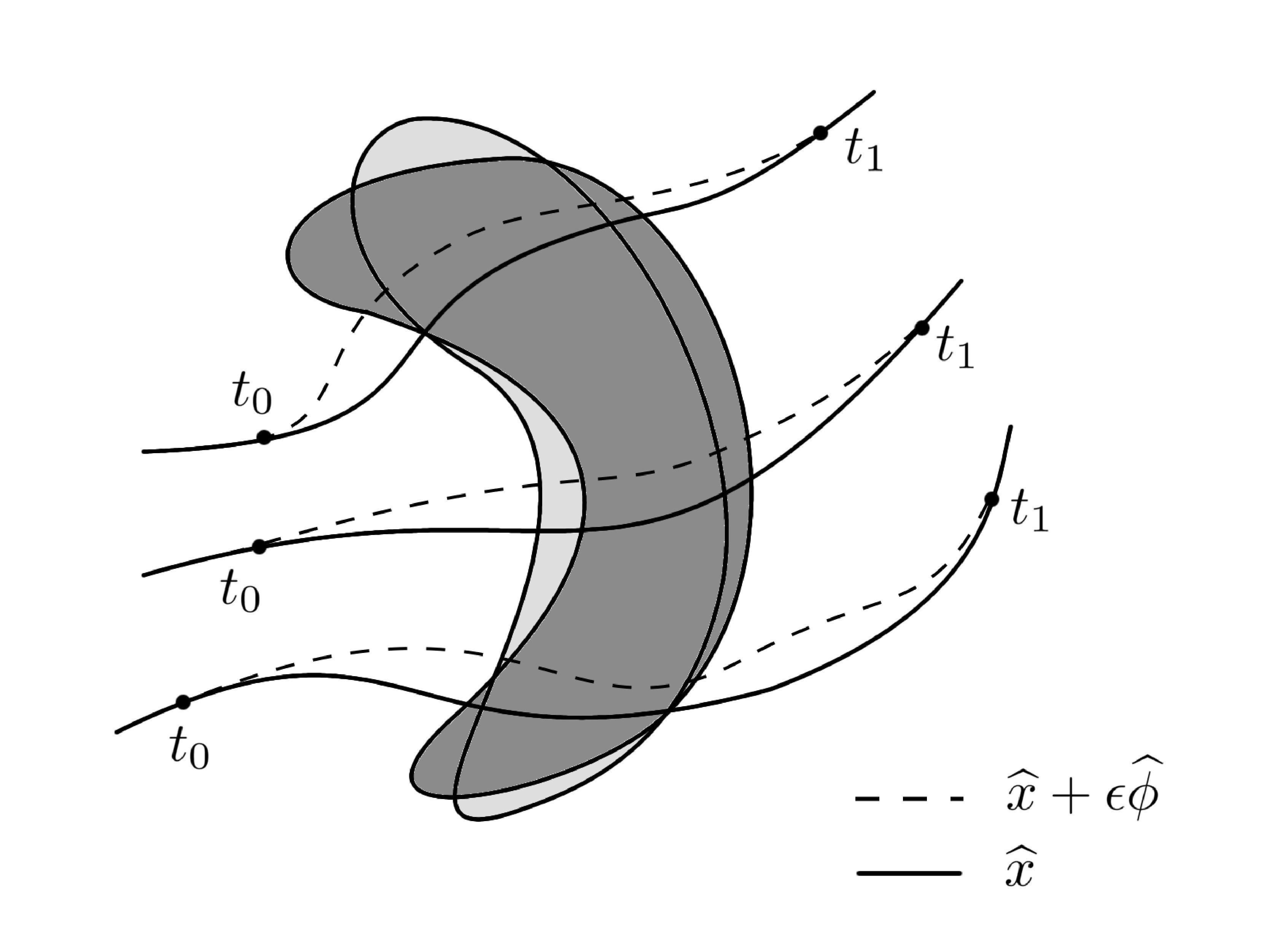}}
\caption{Conjugate flow perturbation $\x^\mu +\epsilon\widehat{\phi}^\mu$ 
induced by a field variation $u^j+\epsilon\varphi^j$ and 
corresponding deformation of the volume of 
particles $\sigma^\nu$ advected by the flow.}
\label{fig:conjugate flow perturbation}
\end{figure}
Note that in these equations we have $\widehat{\phi}^\mu\neq\widetilde{\phi}^\mu$ 
and $\varphi^j\neq \widetilde{\varphi}^j$. In fact, if we arbitrarily 
perform a simultaneous perturbation of $u^j$ and $\x^\mu$ we cannot obviously 
expect that, in general, the functional disturbances arising from the G\^ateaux 
differentials (\ref{conj2}) and (\ref{conj1}) coincide with the perturbations at 
the left hand side of Eqs. (\ref{field_perturbation}) and (\ref{trajectory_perturbation}).
This immediately leads us to the question of which variable between $u^j$ and $\x^\mu$ 
should be chosen as \emph{independent} when performing perturbations. 
In the sequel we will be mostly concerned with perturbations 
induced in the conjugate flow $\x^\mu$ through a variation of 
the solution field $u^j$, i.e. we will mostly employ Eq. (\ref{trajectory_perturbation}), although the other approach, 
i.e. the one based Eq. (\ref{field_perturbation}), can be equivalently
considered.

\section{Conjugate flow representation of field equations}
\label{sec:rep}
Several field equations of mathematical physics, remarkably the fluid 
mechanics equations, include naturally the concept of conjugate flow within 
their formulation. Such flow usually has a direct physical interpretation, e.g., 
trajectories of fluid elements in space, and it often constitutes the 
ground work on which dynamical results are constructed \cite{Aris, Batchelor,Spurk}.
Many other field equations, however, do not include 
explicitly any term having a direct reference to a conjugate flow. 
This is the case, for example, of the classical heat equation, the Maxwell's 
equations of electrodynamics, the laws of elasticity and, undoubtedly, many others.
The fundamental question at this point is whether it is possible to formulate 
a law that include \emph{both} the field equations and the conjugate flow and 
it allows to study their interaction, e.g., in the context of the principle of 
stationary action. The answer is affirmative and the simplest way to 
achieve this result is to represent the field equations relative to 
a coordinate system which is advected by the conjugate flow, namely, 
coordinates $\sigma^\nu$. In other words, we represent the field 
equations on a curvilinear manifold \cite{Calin} which is assumed 
to be functionally dependent on their solution.
As a consequence, the equations look completely different 
in conjugate flow intrinsic coordinates and, in general, they are  
{\em highly nonlinear}. For example, by using the mathematical 
tools summarized in appendix \ref{app:B}, it can be show that the 
classical one-dimensional heat equation 
\begin{equation} 
\frac{\partial U}{\partial t}-\alpha\frac{\partial^2 U}{\partial x^2}=0,\label{onedimfeq}
\end{equation}
where $U(x,t)$ denotes the temperature field in 
fixed Cartesian coordinates, can be written 
in terms of conjugate flow intrinsic coordinates as
\begin{equation}
\frac{\partial u}{\partial t}-\frac{1}{\partial \x/\partial \sigma}
\frac{\partial u}{\partial \sigma}\frac{\partial \x}{\partial t}-
\frac{\alpha}{\left(\partial \x/\partial \sigma\right)^3}
\left(\frac{\partial \x}{\partial \sigma }\frac{\partial^2 u}{\partial \sigma^2}-
\frac{\partial^2 \x}{\partial \sigma^2 }
\frac{\partial u}{\partial \sigma}\right)=0,
\label{ffr}
\end{equation}
where the flow $\x$ is a functional of $u$. For illustration 
purposes, here we have assumed that the time 
variable is not transformed, i.e. we have set $\x^0=\sigma^0=t$. By 
examining Eq. (\ref{ffr}) under the conjugate flow perspective, 
we see that a perturbation in the field $u\left(\sigma,t\right)$ 
induces also a perturbation in the conjugate flow 
$\x\left(\sigma,t;u\right)$ through Eqs. (\ref{trajectory_perturbation}) 
and (\ref{conj1}). 
Therefore the perturbed equation in conjugate flow 
intrinsic coordinates includes many terms arising 
from the perturbations of both $u$ and $\x$. 
Clearly, if the conjugate flow is in rest with respect to the
fixed Cartesian coordinate system then Eq. (\ref{ffr}) coincides 
with Eq. (\ref{onedimfeq}), although the effects of the 
aforementioned functional perturbations are 
still present.

From what has been said, it is clear that the conjugate flow 
representation of a field equation is much 
more complicated than a standard formulation in fixed 
Cartesian coordinates (see table 1). This has been 
observed, e.g., by Temam \cite{Temam1}, in the context of 
the Lagrangian representation of the Navier-Stokes equations.
He pointed out that ``the Lagrangian representation is not used too often 
because the Navier-Stokes equations in Lagrangian coordinates are
highly nonlinear''. Indeed, by using results of appendix \ref{app:B}, 
it can be shown that these equations can be written 
in general conjugate flow intrinsic coordinates as
\begin{equation}
\frac{\partial u^j}{\partial \sigma^\nu}A_0^\nu +
u^k\frac{\partial u^j}{\partial \sigma^\nu}A_k^\nu=-\frac{\partial p}{\partial \sigma^\nu}A_j^\nu+
\frac{1}{\textit{Re}}\left(\frac{\partial^2u^j}{\partial \sigma^\lambda\partial \sigma^\rho}
A^\rho_k A^\lambda_k+\frac{\partial u^j}{\partial \sigma^\lambda}
\frac{\partial A^\lambda_k}{\partial \sigma^\rho}A^\rho_k\right),
\label{NSL}
\end{equation}
where the quantities $A^\mu_\nu$, defined in Eq. (\ref{A}), 
are rather complicated functions of $\x^\mu$. Clearly, when the 
coordinate system $\sigma^\nu$ is advected exactly by 
the physical fluid flow, i.e. when the functional 
link between $u^j$ and $\x^j$ is defined by Eq. (\ref{frel}), 
then Eq. (\ref{NSL}) coincides with the Lagrangian representation of the 
Navier-Stokes equations.

\begin{table}
\label{tab:2}
\begin{tabular}{c|c}
Field Equation & Conjugate Flow Representation\\
\hline\\
$\displaystyle\nabla\cdot\bm U=F$ & 
$\displaystyle\frac{\partial u^j}{\partial \sigma^\nu}A_j^\nu=f$ 
\\\\
$\displaystyle\nabla^2 U=F$ & 
$\displaystyle\left(\frac{\partial^2u}{\partial \sigma^\lambda\partial \sigma^\rho}
A^\rho_k A^\lambda_k+\frac{\partial u}{\partial \sigma^\lambda}
\frac{\partial A^\lambda_k}{\partial \sigma^\rho}A^\rho_k\right)=f$
\\\\
$\displaystyle\frac{\partial U}{\partial t}=
\alpha \nabla^2 U$ &
$\displaystyle\frac{\partial u}{\partial \sigma^\nu}A_{0}^\nu=
\alpha\left(\frac{\partial^2u}{\partial \sigma^\lambda\partial \sigma^\rho}
A^\rho_k A^\lambda_k+\frac{\partial u}{\partial \sigma^\lambda}
\frac{\partial A^\lambda_k}{\partial \sigma^\rho}A^\rho_k\right)$\\
$\displaystyle\frac{\partial U^j}{\partial  t}+\bm U\cdot \nabla U^j=
-\nabla P+\frac{1}{\textit{Re}}\nabla^2 U^j$ &
\begin{minipage}{7cm}
\begin{eqnarray}
\frac{\partial u^j}{\partial \sigma^\nu}A_0^\nu +
u^k\frac{\partial u^j}{\partial \sigma^\nu}A_k^\nu=
-\frac{\partial p}{\partial \sigma^\nu}A_j^\nu+ \nonumber \\
\frac{1}{\textit{Re}}\left(\frac{\partial^2u^j}{\partial \sigma^\lambda\partial \sigma^\rho}
A^\rho_k A^\lambda_k+\frac{\partial u^j}{\partial \sigma^\lambda}
\frac{\partial A^\lambda_k}{\partial \sigma^\rho}A^\rho_k\right)\nonumber
\end{eqnarray}
\end{minipage}\\\\\hline
\end{tabular}
\caption{Conjugate flow representation of well-known 
PDEs of mathematical physics. The quantities $A^\mu_\nu$ are 
rather complicated functions of $\x^\mu$ defined in Eq. (\ref{A}).}
\end{table}

We remark both the conjugate flow 
perturbation as well as the perturbation induced in the 
field equations can be represented relative 
to arbitrary curvilinear coordinates \cite{Venturi_A,Aris,Weinberg}.
The choice of a coordinate system advected by the conjugate 
flow, however, is convenient since the flow map $\x^\mu(\sigma^\nu;u^j)$ 
then appears explicitly in the equations of motion, and it can be 
selected in order to satisfy existence conditions of an action 
functional (see section \ref{sec:action}). 
On the contrary, if we consider fixed Cartesian coordinates then 
the conjugate flow perturbation involves the inverse of the 
flow map $\x^\mu(\sigma^\nu;u^j)$.

\subsection{Functional setting}
\label{sec:Gateaux}
\begin{figure}
\centerline{ \includegraphics[height=8.5cm]{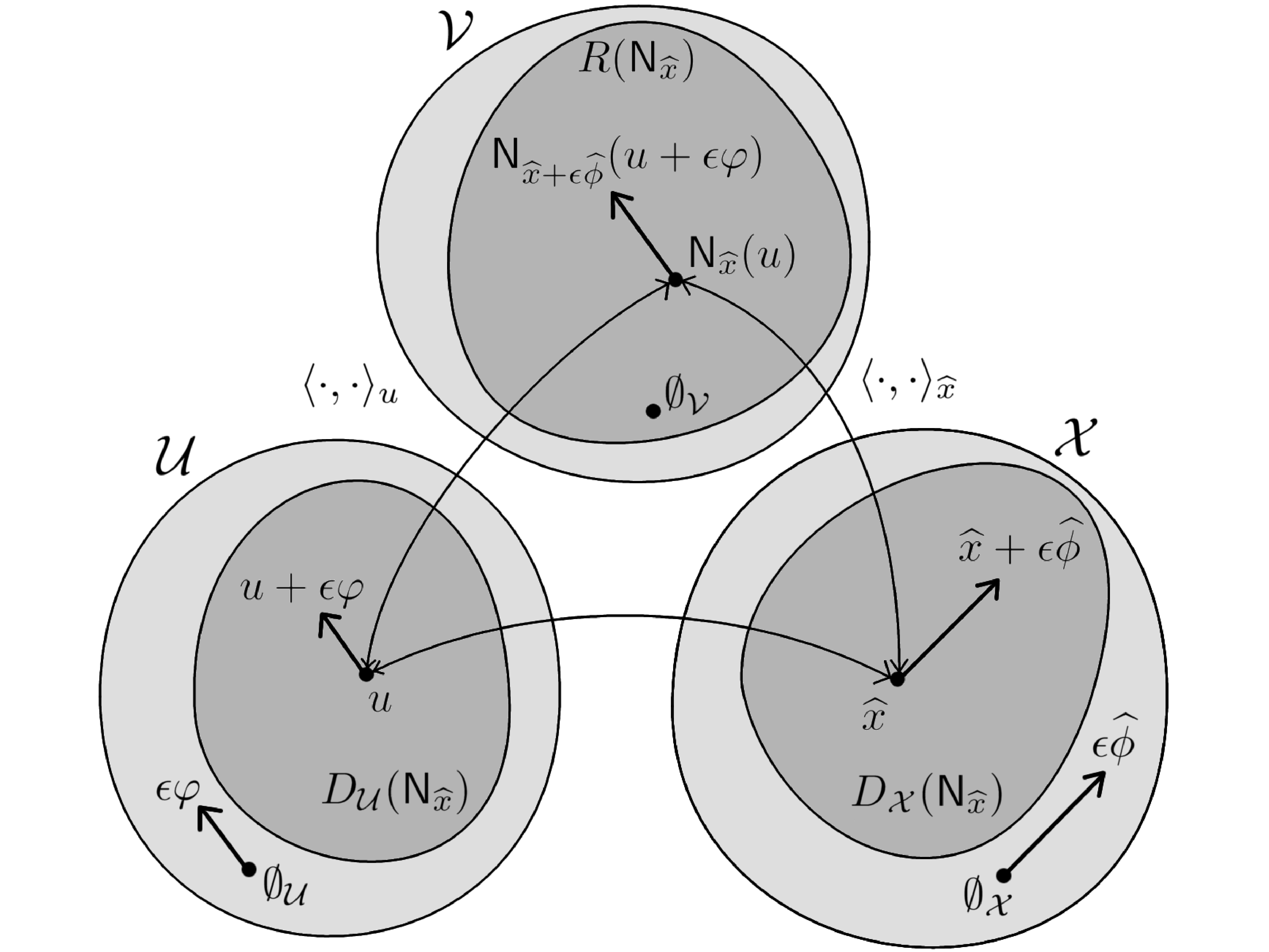}}
\caption{Sketch of the function spaces employed in the functional setting 
of the conjugate flow theory. Shown are the domains $D_\U(\N_{\x})\subseteq \U$
and $D_\X(\N_{\x})\subseteq \X$ of the nonlinear operator $\N_{\x}$ representing 
the field equations. The range of $\N_{\x}$ is denoted by $R(\N_{\x})\subseteq\V$. 
We also show the correspondence between field perturbations $u+\epsilon\varphi$, 
conjugate flow perturbations $\x+\epsilon\widehat{\phi}$ and 
corresponding perturbations induced in the field equations
$\N_{\x+\epsilon\widehat{\phi}}(u+\epsilon\varphi)$ relative to a 
specific representation ($u$, $\x$, $\N_{\x}$). The local bilinear forms that 
put the various spaces in duality are indicated in between the sets.}
\label{fig:duality}
\end{figure}

Let us associate with the physical system the linear function space $\U$, 
whose elements are the $N$-tuples $u=\left(u^1,..,u^N\right)$. 
Similarly, let us consider the \emph{configuration space} $\X$, 
whose elements, denoted as $\x=\left(\x^0,..,\x^n\right)$, represent 
$(n+1)$-dimensional conjugate flows, $n$ being the number of spatial dimensions.
In general, the configuration space is not a linear space because the 
summation of two conjugate flows is not a conjugate flow. This is due 
to the fact that the superimposition of two invertible flows may not be invertible 
(the summation of two invertible Jacobian matrices is not necessarily invertible).
However, the requirement that the perturbed conjugate flow has 
the same properties of the unperturbed one, i.e. that it is still a 
diffeomorphism, is equivalent to state that locally, i.e., in 
the neighborhood of a particular flow $\x$, the configuration 
space $\X$ is linear or can be linearzed. 
This is equivalent to assume that the 
flow perturbation $\widehat{\phi}$ is  
invertible, i.e., that the G\^ateaux differential (\ref{conj1})  
defines an invertible flow map.
In this sense we can say that the configuration 
space $\X$ is \emph{locally linear}.
Given this, an arbitrary field equation (or a system of field equations) 
written in terms of conjugate flow intrinsic coordinates can be synthesized as
\begin{equation}
 \N_{\x}\left(u\right)=\emptyset_\V,
 \label{feq}
\end{equation}
where $\N_{\x}$ is, in general, a nonlinear operator while $\emptyset_\V$ 
denotes the null element of a third topological linear space $\V$.
The subscript $\x$ in $\N_{\x}$ reminds us that the operator is written 
in terms of conjugate flow intrinsic coordinates $\sigma^\nu$, i.e. 
on the manifold defined by $\x$. For example, Eq. (\ref{ffr}) can 
be put in the form (\ref{feq}) by defining $\N_{\x}$ as 
\begin{equation}
\N_{\x}\left(u\right)\Def\frac{\partial u}{\partial t}-\frac{1}{\partial \x/\partial \sigma}
\frac{\partial u}{\partial \sigma}\frac{\partial \x}{\partial t}-
\frac{\alpha}{\left(\partial \x/\partial \sigma\right)^3}\left(\frac{\partial \x}{\partial \sigma }
\frac{\partial^2 u}{\partial \sigma^2}-\frac{\partial^2 \x}{\partial \sigma^2 }
\frac{\partial u}{\partial \sigma}\right).
\end{equation}
The domain of the nonlinear operator $\N_{\x}$ 
is a suitable space of functions satisfying the initial or 
the boundary conditions of the problem. In the conjugate 
flow theory, however, the operator $\N_{\x}(u)$ acts on both $u$ and $\x$ 
and therefore it implicitly identifies two different domains, one within the
space of fields $\U$ and the other one within the configuration space $\X$. 
These two domains will be denoted by $D_\U(\N_{\x})\subseteq \U$ and 
$D_\X(\N_{\x})\subseteq \X$, respectively 
(see figure \ref{fig:duality}). 
The range of the operator $\N_{\x}$ will be denoted by $R(\N_{\x})\subseteq \V$.
The next fundamental step in the functional setting of the 
conjugate flow theory is to introduce duality 
parings between the linear spaces $\U$, $\V$ and the locally linear one $\X$ through 
non-degenerate \emph{local bilinear forms} \cite{Venturi2,Magri}. 
To this end, let us define
\begin{eqnarray}
\langle \cdot , \cdot\rangle_u&:&\V\times \U \rightarrow \mathbb{R},\label{Local1}\\
\langle \cdot, \cdot\rangle_{\x}&:&\V\times \X \rightarrow \mathbb{R}.\label{Local2}
\end{eqnarray}
The subscripts $u$ and $\x$ in Eqs. (\ref{Local1}) and (\ref{Local2}) 
emphasize the fact that such forms depend 
also on $u$ and $\x$, in a possibly 
nonlinear way. An explicit expression of (\ref{Local1}) will be given in 
section \ref{sec:classical action}. The 
forms (\ref{Local1}) and (\ref{Local2}) can be put in a 
correspondence through the linear transformations 
defined by Eqs. (\ref{conj2}) and (\ref{conj1}). In fact, as shown in 
figure \ref{fig:duality}, the elements of $D_\U(\N_{\x})$ in the 
neighborhood of a certain $u$ are in correspondence with 
the elements of $D_\X(\N_{\x})$ in the neighborhood of a certain $\x$. 
In practice, such correspondence can be established locally 
through the linear operators $\delta u/ \delta \x$ and $\delta \x/ \delta u$. 
For instance, by using Eq. (\ref{conj2}) we obtain
\begin{equation}
\langle v,\widetilde{\varphi}\rangle_u =\langle v,\frac{\delta u}{\delta \x}\widetilde{\phi}\rangle_u 
=\langle v ,\widetilde{\phi}\rangle_{\x}.
\end{equation}
We shall conclude this section by explaining why we have chosen the definition 
``conjugate flow'' for the transformation (\ref{tr_conj}). 
To this end, let us recall that the G\^ateaux differential 
of $\x$ with respect to $u$ defines a linear functional from 
the space $\U$ to the space $\X\equiv \U^\dagger$, which 
is the \emph{conjugate} space
of $\U$. Thus, for every admissible $u\in \U$, the flow $\x$ belongs 
to the conjugate space of $\U$, hence the definition ``conjugate flow''.
In a broader sense, the adjective ``conjugate'' simply emphasizes 
that there exists a functional relation between the flow $\x$, 
the governing  equations of the field theory, $\N_{\x}(u)=\emptyset_\V$, 
and the solution field $u$. 
We also remark that a definition of conjugate flow already 
appeared in the literature \cite{Benjamin,Benjamin1}, as
``a flow uniform in the direction of streaming which separately 
satisfy the hydrodynamical equations''. Such definition is of course
very different from ours.

\subsection{Perturbation expansions}
By using the general operator framework developed in the previous 
section, we can easily synthesize in a single operator equation the perturbative 
form of an arbitrary set of field equations 
in the presence of a conjugate flow perturbation, i.e. a simultaneous 
perturbation of both the solution field and the conjugate flow. 
Indeed, to the first-order in $\epsilon$ we have
\begin{equation}
\N_{\x+\epsilon\widehat{\phi}}\left(u+\epsilon\varphi\right)=\N_{\x}\left(u\right)+
\epsilon\left[\frac{\delta \N_{\x}}{\delta u}\varphi+\frac{\delta \N_{\x}}{\delta \x}\widehat{\phi}\right]
+\cdots ,\label{expansion2}
\end{equation}
where the G\^ateaux differentials appearing in Eq. (\ref{expansion2}) are defined as
\begin{eqnarray}
\frac{\delta \N_{\x}}{\delta u}\varphi&\Def&\lim_{\epsilon\rightarrow0}
\frac{\N_{\x}\left(u+\epsilon\varphi\right)-\N_{\x}\left(u\right)} {\epsilon},\label{g1}\\
\frac{\delta \N_{\x}}{\delta \x}\widehat{\phi}&\Def&\lim_{\epsilon\rightarrow0}
\frac{\N_{\x+\epsilon\widehat{\phi}}\left(u\right)-\N_{\x}\left(u\right)} {\epsilon},
\end{eqnarray}
provided that such limits exist.
A function space representation of the conjugate flow perturbation 
is sketched in figure \ref{fig:duality}.

\section{Conjugate flow action functionals}
\label{sec:action}
Let us consider a field $u\in D_\U(\N_{\x})$ and a conjugate 
flow $\x\in D_\X(\N_{\x})$.
The couple $(u,\x)$ does not necessarily have to be a 
solution to the field equation, i.e. $\N_{\x}(u)\neq \emptyset_\V$.
Disregarding the particular form of the operator $\N_{\x}$, 
it is useful to consider
\begin{equation}
v=\N_{\x}\left(u\right)\in R(\N_{\x})\label{h1}
\end{equation}
as a definition two vector fields \cite{Nashed}: one in $D_\U(\N_{\x})$ 
and the other one in $D_\X(\N_{\x})$, respectively. This allows us to 
introduce in a conceptually simple way the notion of a line 
integral of an operator according to a geometric standpoint which 
seems originally due to Volterra \cite{Volterra1}.
To this end, let us consider a one-parameter family of fields
in the domain $D_\U(\N_{\x})$
\begin{equation}
 u=u_\lambda\quad\quad\left(0\leq\lambda\leq 1\right).
\end{equation}
This can be regarded as a line in the function space $D_\U(\N_{\x})$.
With such line we can associate the number
\begin{equation}
\ell_u=\int_{0}^{1}\langle \N_{\x}\left(u_\lambda\right),
\frac{\partial u_{\lambda}}{\partial \lambda}\rangle_{u_\lambda} d\lambda,
\label{circulation}
\end{equation}
i.e. the path integral of the operator $\N_{\x}$ along the trajectory of 
functions $u_\lambda\in D_\U(\N_{\x})$. 
We recall that $\langle\cdot,\cdot\rangle_u$ 
in (\ref{circulation}) denotes the local bilinear form (\ref{Local1}). 
In the context of the conjugate flow theory, we can also 
define the path integral of the operator $\N_{\x}$ 
along a trajectory of flows $\x_\lambda$ in the 
space $D_\X(\N_{\x})$, i.e. 
\begin{equation}
\ell_{\x}=\int_{0}^{1}\langle \N_{\x_\lambda}\left(u\right),
\frac{\partial \x_{\lambda}}{\partial \lambda}\rangle_{\x_\lambda} d\lambda,
\label{circulation1}
\end{equation}
where  $\langle\cdot,\cdot\rangle_{\x}$ denotes the local 
bilinear form (\ref{Local2}).
If the line integrals (\ref{circulation}) and (\ref{circulation1}) are 
independent of the path of integration then the operator $\N_{\x}$ 
is said to be {\em potential} with respect to the chosen 
local bilinear form.
In this case the integral from a prefixed element $u_0$ 
to any element $u$ in $D_\U(\N_{\x})$ along an arbitrarily 
chosen path defines the {\em action functional}
\begin{equation}
A_u\left[u\right]= A_u\left[u_0\right]+\int_{0}^{1}\langle \N_{\x}\left(u_\lambda\right),
\frac{\partial u}{\partial \lambda}\rangle_{u_\lambda} d\lambda.
\label{the potential}
\end{equation} 
Similarly, the line integral from a prefixed conjugate flow $\x_0$ to another
flow $\x$ along an arbitrarily chosen line in $D_\X(\N_{\x})$ 
defines another (dual) action functional
\begin{equation}
A_{\x}\left[\x\right]= A_{\x}\left[\x_0\right]+\int_{0}^{1}\langle \N_{\x_{\lambda}}\left(u\right),
\frac{\partial \x_\lambda}{\partial \lambda}\rangle_{\x_\lambda} d\lambda.
\label{the potential1}
\end{equation}
In turn, the operator $\N_{\x}$ is said to be 
the {\em gradient} of the functionals $A_u[u]$ or $A_{\x}[\x]$. 
This definition relies on the fact that if we calculate the 
infinitesimal variation of (\ref{the potential}) and (\ref{the potential1}) 
with respect to independent variations of $u$ and $\x$, respectively,
then we obtain
\begin{eqnarray}
\delta A_u\left[u\right]=\langle \N_{\x}\left(u\right), \delta u\rangle_u,\quad\quad
\delta A_{\x}\left[\x\right]=\langle \N_{\x}\left(u\right), \delta \x\rangle_{\x}.
\label{areyousure1}
\end{eqnarray}
These relations show that the 
equations of motion of the system, i.e. $\N_{\x}\left(u\right)=\emptyset_\V$, 
can be obtained as a stationary point of either $A_u[u]$ or $A_{\x}[\x]$,
for arbitrary variations $\delta u$ and $\delta \x$, respectively.
Thus, the theory of conjugate flows allows us to look for 
action functionals associated with field equations in two 
different ways, depending on which 
variable between $u$ or $\x$ is assumed 
as \emph{independent}. Clearly, if we consider $u$ as 
independent then we are looking for the set of conjugate flows such that the 
field equation is potential. On the contrary, if we consider the conjugate 
flow $\x$ as independent then we are looking for the set of fields $u$ 
such that the field equation is potential.
In the sequel we will be mostly concerned with conjugate flows 
action functionals where the field $u$ is considered as 
independent variable.

\subsection{Existence conditions}
\label{sec:existence}
In order to formulate the existence conditions of conjugate flow 
action functionals we follow the approach of Magri \cite{Magri}. 
To this end, we consider two infinitesimal trajectories 
(two infinitesimal straight lines) of the field $u$ in 
the function space $D_\U(\N_{\x})$
\begin{eqnarray}
\textrm{I}&:& u\rightarrow u+\epsilon\varphi\nonumber\\
\textrm{II}&:& u\rightarrow u+\nu\psi\nonumber
\end{eqnarray}
Correspondingly, we have the following infinitesimal conjugate flow perturbations 
\begin{eqnarray}
\textrm{I}&:& \x\rightarrow\x+\epsilon\widehat{\phi}\nonumber\\
\textrm{II}&:& \x\rightarrow\x+\nu\widehat{\eta}\nonumber
\end{eqnarray}
where $\widehat{\phi}$ and $\widehat{\eta}$ are related to the field perturbations 
$\varphi$ and $\psi$ by Eq. (\ref{conj1}). Due to this fundamental relation, an infinitesimal circulation of the 
operator $\N_{\x}$ around the element $u\in D_\U(\N_{\x})$ induces 
an infinitesimal circulation of $\N_{\x}$ around a flow 
$\x\in D_\X(\N_{\x})$. The vanishing of these simultaneous circulations with 
respect to the local bilinear form (\ref{Local1}) is synthesized by the condition
\begin{eqnarray}
\langle \N_{\x}\left(u\right),\epsilon\varphi\rangle_{u}+\langle 
\N_{\x+\epsilon\widehat{\phi}}\left(u+\epsilon\varphi\right),\nu\psi\rangle_{u+\epsilon\varphi}= \nonumber\\ 
\langle \N_{\x}\left(u\right),\nu\psi\rangle_{u}+\langle \N_{\x+\nu\widehat{\eta}}\left(u+\nu\psi\right),
\epsilon\varphi\rangle_{u+\nu\psi}.
\label{curl}
\end{eqnarray}
To the second-order in $\epsilon$ and $\nu$ we have
\begin{eqnarray}
\langle \N_{\x+\epsilon\widehat{\phi}}\left(u+\epsilon\varphi\right),\nu\psi \rangle_{u+\epsilon\varphi}
=\langle \N_{\x}\left(u\right),\psi\rangle_{u}+\nonumber\\
\epsilon\nu\langle\frac{\delta \N_{\x}}{\delta u}\varphi+\frac{\delta \N_{\x}}{\delta \x}\widehat{\phi},
\psi \rangle_{u}+\epsilon\nu\langle\varphi;\N_{\x}(u),\psi\rangle_u,
\label{relation}
\end{eqnarray}
where
\begin{equation}
\langle\varphi;v,\psi\rangle_u\Def \lim_{\epsilon\rightarrow0}\frac{\langle v,\psi\rangle_{u+\epsilon\varphi}-\langle v,\psi\rangle_{u}}{\epsilon}
\end{equation}
denotes the G\^ateaux differential of the local bilinear form (\ref{Local1}), considered as a 
particular type of nonlinear operator on $u$. A substitution of Eq. (\ref{relation}) 
into Eq. (\ref{curl}) gives
\begin{eqnarray}
\langle\frac{\delta \N_{\x}}{\delta u}\varphi+\frac{\delta \N_{\x}}{\delta \x}\widehat{\phi},\psi \rangle_{u}+\langle\varphi;\N_{\x}(u),\psi\rangle_u=\nonumber\\
\langle\frac{\delta \N_{\x}}{\delta u}\psi+\frac{\delta \N_{\x}}{\delta \x}\widehat{\eta},\varphi \rangle_{u}+\langle\psi;\N_{\x}(u),\varphi\rangle_u.
\label{curl1}
\end{eqnarray}
Finally, by using Eq. (\ref{conj1}) we can write 
the vanishing condition of the infinitesimal circulation entirely 
in terms of field perturbations $\psi$ and $\varphi$ as
\begin{eqnarray}
\langle\G_{\x}\varphi,\psi \rangle_{u}+\langle\varphi;\N_{\x}(u),\psi\rangle_u=\langle\G_{\x}\psi,\varphi \rangle_{u}+\langle\psi;\N_{\x}(u),\varphi\rangle_u,
\label{curl4}
\end{eqnarray}
where the linear operator $\G_{\x }$ is defined as
\begin{equation}
\G_{\x }\Def\frac{\delta \N_{\x}}{\delta u}+\frac{\delta \N_{\x}}{\delta \x }\frac{\delta \x }{\delta u}\label{G}.
\end{equation}
Thus, if the circulation vanishes along any infinitesimal closed line 
in $D_\U(\N_{\x})$ then Eq. (\ref{curl4}) must hold for 
every $\varphi$, $\psi$ and for all admissible $u$. This is the necessary 
condition for operators to be potential  
with respect to the local bilinear form (\ref{Local1}). If the domain of 
the operator $\N_{\x}$ is simply connected, then this condition is 
also sufficient. This happens, e.g., when $D_\U(\N_{\x})$  
is defined by linear homogeneous initial or boundary 
conditions (in this case $D_\U(\N_{\x})$ is a convex set). 
We remark that the general symmetry condition (\ref{curl4}) includes interesting subcases. For example, if the flow $\x$ is not a functional of $u$ then we have 
\begin{equation}
\langle\frac{\delta \N_{\x}}{\delta u}\varphi,\psi \rangle_{u}+
\langle\varphi;\N_{\x}(u),\psi\rangle_u=
\langle\frac{\delta \N_{\x}}{\delta u}\psi,\varphi \rangle_{u}+
\langle\psi;\N_{\x}(u),\varphi\rangle_u.
\label{curl6}
\end{equation}
In addition, if the bilinear form $\langle\cdot,\cdot\rangle_u$ 
does not depend on $u$, i.e., if we are dealing with a standard (non-local) 
bilinear form, then the symmetry condition (\ref{curl6}) 
coincides with the classical one obtained by Vainberg \cite{Vainberg}
\begin{equation}
\langle\frac{\delta \N_{\x}}{\delta u}\varphi,\psi \rangle=\langle\frac{\delta \N_{\x}}{\delta u}\psi,\varphi \rangle,
\label{curl2}
\end{equation}
namely, the G\^ateaux derivative of the operator $\N_{\x}$ must be symmetric 
with respect to the bilinear form $\langle\cdot,\cdot\rangle$.

\subsection{Choice of the local bilinear form}
\label{sec:classical action}
In classical, relativistic and quantum field theories the action 
functional has the standard form \cite{Landau,Lovelock,Weinberg}
\begin{equation}
A=\int\mathcal{L}\sqrt{g}d^4\eta.
\label{Ac}
\end{equation}
where $\mathcal{L}$ denotes the Lagrangian density, $g$ is the determinant of the metric 
tensor associated with the coordinate system $\eta$ and $\sqrt{g}d^4\eta$ is the invariant 
space-time volume element ($d^4\eta$ being a shorthand notation for 
$d\eta^0d\eta^1\cdots d\eta^3$). 
A comparison between Eq. (\ref{Ac}) and Eq. (\ref{the potential}) 
suggests that the local bilinear form to be considered for the 
conjugate flow formulation of the inverse problem of the calculus 
of variations is
\begin{equation}
 \langle a,b\rangle_{u}\Def\int_{\Sigma} ab Jd^4\sigma,\quad\quad a\in\U,
 \quad b\in\V,
 \label{local_bilinear_form}
\end{equation}
where the Jacobian determinant $J$ is a rather complicated function 
of $\x$ (see Eq. (\ref{J})). The domain $\Sigma$ appearing in the integral 
(\ref{local_bilinear_form}) is a four-dimensional volume of particles $\sigma^\nu$
advected by the conjugate flow (see figure \ref{fig:two_conj_flows}). 
Before proceeding further, it is useful to clarify the physical 
meaning of the bilinear form (\ref{local_bilinear_form}). 
To this end, with reference to figure \ref{fig:two_conj_flows}, 
let $V(t;u)$ be the volume of particles 
obtained by advecting $\Sigma$ with the conjugate 
flow $\x^\mu(\sigma^\nu;u)$. In particular, 
in figure \ref{fig:two_conj_flows} we show two different 
volumes at time $t_1$, corresponding to two realizations 
of the flow, namely, $\x^\mu(\sigma^\nu;u_1)$ 
and $\x^\mu(\sigma^\nu;u_2)$. 
The bilinear form (\ref{local_bilinear_form}) is defined 
by integration over a volume of particles co-moving 
with the conjugate flow. In fact, 
\begin{equation}
\int_{V(t;u)} A(x^\mu)B(x^\mu)d^4x = 
\int_{\Sigma} a(\sigma^\nu)b(\sigma^\nu)
J(\x^\mu(\sigma^\nu;u))d^4\sigma.
\label{example}
\end{equation}
At the left hand side of this equation we are using fixed Cartesian 
coordinates $x^\mu$, while at the right hand side we are 
using conjugate flow intrinsic coordinates $\sigma^\nu$
($J$ is the Jacobian determinant of the transformation from fixed Cartesian 
to conjugate flow intrinsic coordinates). At each specific time, e.g., 
at time $t_1$ in figure \ref{fig:two_conj_flows}, the trajectory of 
fields $u_\lambda=(1-\lambda) u_1 +\lambda u_2$
($0\leq \lambda\leq 1$) defines a transformation 
that takes the volume $V(t;u_1)$ into the volume $V(t;u_2)$.

The bilinear form (\ref{local_bilinear_form}) generalizes the 
one appearing in the Herivel-Lin variational 
principle \cite{Herivel,Bretherton,Eckart,Serrin}, where the volume 
of particles is advected precisely by the physical fluid flow.
Note also that (\ref{local_bilinear_form}) is symmetric, non-degenerate 
and non-negative, i.e., it satisfies all the properties of an inner product.
By using Eq. (\ref{perturbedJ}) we obtain the following G\^{a}teaux 
derivative
\begin{eqnarray}
\langle \varphi;a,b\rangle_u&\Def&\frac{d}{d\epsilon}  \left[\langle a,b\rangle_{u+\epsilon\varphi} \right]_{\epsilon=0}\nonumber\\
&=&\int_{\Sigma} ab J \nabla \cdot\widehat{\phi}d^4\sigma\label{gd}
\end{eqnarray}
where, according to Eq. (\ref{conj1}), $\widehat{\phi}$ is a linear functional 
of $\varphi$, i.e. (\ref{gd}) is a trilinear form in $a$, $b$ and $\varphi$.
A substitution of Eqs. (\ref{local_bilinear_form}) and (\ref{gd}) into 
the symmetry condition (\ref{curl4}) suggests that for {\em incompressible}
flow perturbations ($\nabla \cdot\widehat{\phi}=0$) we have
\begin{equation}
\langle\G_{\x}\varphi,\psi \rangle_{u}=\langle\G_{\x}\psi,\varphi \rangle_{u}.
\label{er}
\end{equation}
Equation (\ref{er}) requires the symmetry of the operator $\G_{\x }$ 
defined in (\ref{G}) relative to the local inner 
product (\ref{local_bilinear_form}). Thus, the application of the conjugate flow theory to the inverse problem of the 
calculus of variations is now reduced to look for an incompressible 
four-dimensional flow symmetrizing the operator $\G_{\x}$ 
with respect to the bilinear form  (\ref{local_bilinear_form}).

\subsection{Conservation laws}

Once a conjugate flow satisfying the symmetry 
conditions (\ref{curl4}) or (\ref{er}) has been identified, 
we can write the action functional (\ref{the potential}) 
entirely in terms of the field $u$. In particular, 
if we consider the local bilinear form (\ref{local_bilinear_form}) 
we obtain
\begin{equation}
 A_u\left[u\right] = \int_{0}^{1}\int_{\Sigma}\N_{\x(u_{\lambda})}\left(u_\lambda\right)
\frac{\partial u_\lambda}{\partial \lambda}J(u_\lambda) d^4\sigma d\lambda, \label{Afunct}
\end{equation}
where we have set $A\left[u_0\right]=0$. 
A comparison between (\ref{Afunct})
and (\ref{Ac}) yields the following 
{\em Lagrangian density}
\begin{equation}
 \mathcal{L} = \int_{0}^{1} \N_{\x(u_{\lambda})}\left(u_\lambda\right)J(u_\lambda)
\frac{\partial u_\lambda}{\partial \lambda} d\lambda,\label{lagrangian}
\end{equation}
which is expressed in conjugate flow intrinsic coordinates $\sigma^\nu$. 
Such representation can be of course transformed to arbitrary coordinates, 
leading us to the general form (\ref{Ac}). 
Before concluding this section, let us briefly recall that 
the symmetries of the action functional (\ref{Afunct}) are 
associated with conservation laws. The classical approach to 
perform this type of analysis relies on the identification 
of suitable transformation groups (symmetry groups) leaving the 
action invariant \cite{Lovelock,Gelfand,Bluman}.
Therefore, once the conjugate flow action functional is available, 
we can use a fairly well developed framework to compute the N\"oether 
invariants of the field theory and the associated currents. 
This could lead to the discovery of new conservation laws 
in fluid dynamics and other disciplines.

\vspace{1.cm}

\section{Symmetrizing flows}
\label{sec:formal symmetry}
A field equation is said to be formally symmetric when the 
operator symmetry condition, e.g. Eq. (\ref{er}), is satisfied 
disregarding the particular form of the boundary or the initial 
conditions associated with the problem. Clearly, when the domain of 
the operator $\N_{\x}$ is formed by a set of functions satisfying local 
homogeneous boundary and initial conditions then formal symmetry is a necessary
condition for symmetry. Such a condition, however, is {\em not} sufficient even 
in the case of homogeneous boundaries \cite{Filippov}.
In any case, it is useful to establish formal symmetry conditions for particular 
classes of field equations. This has been done, e.g., by Tonti \cite{Tonti1,Tonti2}
by using classical inner products in fixed coordinate systems.
In this section we obtain similar conditions for 
incompressible conjugate flow variations. 
In particular, we study scalar and vector 
field theories governed by second-order nonlinear 
partial differential equations. These theories include many 
well-known equations of mathematical physics, e.g., 
the Navier-Stokes equations, the Maxwell's equations in potential 
formulation, the laws of elasticity, the advection-reaction-diffusion 
equations and the Schr\"odinger equation.
\subsection{Vector field theories}
The conjugate flow representation of second-order vector field equations 
can be written in the general form
\begin{equation}
\N_{\x}(u)=f_k\left(u^j; u^j_{,\mu}; u^j_{,\mu\nu}; 
\x^\mu_{,\nu}; \x^\mu_{,\nu\lambda} \right)=0,
\label{opeq}
\end{equation}
where the comma denotes partial differentiation with respect to 
$\sigma^\mu$ ($\mu=0,..,3$), i.e., $u^j_{,\mu}\Def\partial u^j/\partial \sigma^\mu$. 
An example of (\ref{opeq}) is the intrinsic form of the 
Navier-Stokes equations (\ref{NSL}), together
with the divergence-free condition of the velocity field.
The G\^ateaux differential of (\ref{opeq}), for arbitrary variations 
of $u^j$ and $\x^\mu$, reads
\begin{eqnarray}
\frac{\delta \N_{\x}}{\delta u}\varphi+\frac{\delta \N_{\x}}{\delta \x}\widehat{\phi}=
\frac{\partial f_k }{\partial u^j}\varphi^j+
\frac{\partial f_k }{\partial u^j_{,\mu}}\varphi^j_{,\mu}+
\frac{\partial f_k}{\partial u^j_{,\mu\nu}}\varphi^j_{,\mu\nu}+
\frac{\partial f_k}{\partial \x^\mu_{,\nu}}\widehat{\phi}^\mu_{,\nu}+
\frac{\partial f_k }{\partial \x^\mu_{,\nu\lambda}}\widehat{\phi}^\mu_{,\nu\lambda}.
\label{G0}
\end{eqnarray}
At this point we recall that the conjugate flow 
perturbation $\widehat{\phi}^\mu$ is related to the field 
perturbation $\varphi^j$ by Eq. (\ref{conj1}).  In general, such 
relation involves both derivatives and integrals. 
For example, it could be in the form
\begin{equation}
\widehat{\phi}^\mu=\int_{\Sigma} K_j^\mu(\sigma^\nu;u^k)\varphi^j d^4\sigma + 
W_j^\mu\left(\sigma^\nu;u^k\right)\varphi^j+
M^\mu_{j\lambda}\left(\sigma^\nu;u^k\right)\varphi^j_{,\lambda},
\label{form 1}
\end{equation}
where $\Sigma$ is a four-dimensional volume of particles 
advected by the conjugate flow and $K_j^\mu$, $W_j^\mu$,  
$M^\mu_{j\lambda}$ are suitable functions.

If we proceed in the most general case, i.e., by using
Eq. (\ref{conj1}) without any further specification, then 
the equations for the conjugate flow satisfying the
symmetry condition (\ref{er}) will be {\em functional 
differential equations}.
In practical application, however, it might be convenient 
to restrict the type of functional dependence to 
specific forms, for example (\ref{form 1}).
In these cases, the type of functional dependence between $\widehat{\phi}^\mu$ 
and $\varphi^j$ is somehow imposed and it defines a {\em class} 
of conjugate flows. If such class is large enough, then the symmetry 
condition (\ref{er}) will extract one or more flows 
for which the vector field theory is potential.
Let us illustrate this procedure with reference
to incompressible conjugate flows that are {\em local functionals} of the 
solution field $u^j$. In particular, we shall consider
\begin{equation}
 x^\mu=\x^\mu\left(\sigma^\nu,u^j\right),\label{soe}
\end{equation}
where $\x^\mu$ (to be determined) are {\em algebraic functions} 
of $u^j$ which do not involve integrals in space-time, e.g., $\x^\mu$ 
are polynomials of $u^j$. 
Note that the assumption that $\x^\mu$ are local functionals 
of $u^j$ could lead to determining equations for the conjugate flow 
with no solutions. In other words, it might be possible that 
the class of algebraic conjugate flows is not large enough
to accommodate the symmetry requirement (\ref{er}) for a 
specific set of PDEs. In any case, it is interesting to analyze such 
class and derive evolution equations for the symmetrizing flow.
To this end, let us first set 
\begin{eqnarray}
a^\mu_j&\Def&\frac{\partial \x^\mu}{\partial u^j},\label{56}\\
b^\mu_{j\nu}&\Def&a^\mu_{j,\nu}+\frac{\partial a_j^\mu}{\partial u^k}u^k_{,\nu},\label{bmu}\\
c^\mu_{j\nu\rho}&\Def&b^\mu_{j\nu,\rho}+\frac{\partial b^\mu_{j\nu}}{\partial u^i}u^i_{,\rho}\label{58}.
\end{eqnarray}
This allows us to write the conjugate flow perturbation $\widehat{\phi}^\mu$ 
and its partial derivatives as
\begin{eqnarray}
\widehat{\phi}^\mu&=&a_j^\mu\varphi^j,\\
\widehat{\phi}^\mu_{,\nu}&=&b^\mu_{j\nu}\varphi^j+a^\mu_j\varphi^j_{,\nu},\label{firstder}\\
\widehat{\phi}^\mu_{,\nu\rho}&=&c^\mu_{j\nu\rho}\varphi^j+b^\mu_{j\nu}\varphi^j_{,\rho}+
b^\mu_{j\rho}\varphi^j_{,\nu}+a_j^\mu\varphi^j_{,\nu\rho}.\label{secondder}
\end{eqnarray}
Remarkably, the highest derivative order 
in $\widehat{\phi}^\mu$ and $\varphi^j$ is the same. 
In fact, the $k$th-order derivative of $u^j$ with respect 
to $\sigma^\nu$ involves the $k$th-order derivative 
of $\x^\mu$ (see Eqs. (\ref{covariant}) and (\ref{second})). This is why 
we have included the second-order derivative of the conjugate flow in the 
second-order vector field equation (\ref{opeq}). 
A substitution of (\ref{firstder})-(\ref{secondder}) into (\ref{G0}) 
yields the following operator $\G_{\x}$ (see Eq. (\ref{G}))
\begin{equation}
 \G_{\x} \varphi = H_{kj}\varphi+B_{kj}^\nu \varphi_{,\nu}^j+F^{\mu\nu}_{kj}\varphi^j_{,\mu\nu},
 \label{GG}
\end{equation}
where
\begin{eqnarray}
H_{kj}&\Def&\frac{\partial f_k}{\partial u^j}+
\frac{\partial f_k}{\partial \x^\mu_{,\nu}}b^\mu_{j\nu}+
\frac{\partial f_k}{\partial \x^\mu_{,\nu\rho}}c^\mu_{j\nu\rho},\\
B^\nu_{kj}&\Def& \frac{\partial f_k}{\partial u^j_{,\nu}}+
\frac{\partial f_k}{\partial \x^\mu_{,\nu}}a_j^\mu+
\left(\frac{\partial f_k}{\partial \x^\mu_{,\nu\rho}}+
\frac{\partial f}{\partial \x^\mu_{,\rho\nu}}\right)b^\mu_{j\rho},\label{B}\\
F^{\mu\nu}_{kj}&\Def&\frac{\partial f_k}{\partial u_{,\nu\mu}}+
\frac{\partial f_k}{\partial \x^\rho_{,\mu\nu}}a^\rho_j.\label{F}
\end{eqnarray}
Note that $F^{\mu\nu}_{kj}$ is symmetric with respect to $\mu$ and $\nu$ by 
construction. At this point, we use the local inner product (\ref{local_bilinear_form}) 
and write explicitly the symmetry condition (\ref{er}) as
\begin{eqnarray}
\int_{\Sigma} \psi^k H_{kj}\varphi^j Jd^4\sigma+ 
\int_{\Sigma} \psi^k B_{kj}^\mu\varphi^j_{,\mu} Jd^4\sigma  +
\int_{\Sigma} \psi^k F_{kj}^{\mu\nu}\varphi^j_{,\mu\nu} Jd^4\sigma=\nonumber\\
\int_{\Sigma} \varphi^j H_{jk} \psi^k Jd^4\sigma+ 
\int_{\Sigma} \varphi^j B_{jk}^\mu\psi^k_{,\mu} Jd^4\sigma  +
\int_{\Sigma} \varphi^j F^{\mu\nu}_{jk}\psi^k_{,\mu\nu} Jd^4\sigma.
\label{symmetry1}
\end{eqnarray}
The next step is to reduce the derivative order of the field 
$\psi^k$ at the right hand side by using integration by parts. 
Since we are looking for a formal symmetry condition, we neglect 
all the boundary terms. This yields, for instance
\begin{eqnarray}
\int_{\Sigma} \varphi^j B_{jk}^\mu\psi^k_{,\mu} Jd^4\sigma&=&
-\int_{\Sigma}\left( \varphi^j B_{jk}^\mu J\right)_{,\mu}\psi^kd^4\sigma\nonumber\\
&=&
-\int_{\Sigma}\psi^k B_{jk}^\mu\varphi^j_{,\mu}Jd^4\sigma
-\int_{\Sigma}\psi^k B^\mu_{jk,\mu}\varphi^j Jd^4\sigma-
\int_{\Sigma}\psi^k B_{jk}^\mu \Gamma^\lambda_{\lambda\mu}\varphi^j Jd^4\sigma,
\label{byparts1}
\end{eqnarray}
where the derivative of the Jacobian $J$ has been represented 
in terms of the affine connection (\ref{affine}) 
by using Eq. (\ref{derJac}). 
Similarly, the integral involving $\psi^k_{,\rho\nu}$ in (\ref{symmetry1})
can be rewritten as
\begin{eqnarray}
\int_{\Sigma} \varphi^j F^{\mu\nu}_{jk}\psi^k_{,\mu\nu} Jd^4\sigma&=& 
\int_{\Sigma} \left(\varphi^j F^{\rho\nu}_{jk} J\right)_{,\mu\nu}\psi^k d^4\sigma\nonumber \\
&=&\int_{\Sigma}\left[\varphi^j_{,\mu\nu}F^{\mu\nu}_{jk}  +\varphi^j_{,\mu}F^{\mu\nu}_{jk,\nu} +
\varphi^j_{,\mu}F^{\mu\nu}_{jk}\Gamma^\lambda_{\lambda\nu}  
+\varphi^j_{,\nu}F^{\mu\nu}_{jk,\mu} +\varphi^jF^{\mu\nu}_{jk,\mu\nu} +\right.\nonumber\\
& &\qquad \varphi^jF^{\mu\nu}_{jk,\mu} \Gamma^\lambda_{\lambda\nu}+
\varphi^j_{,\nu}F^{\mu\nu}_{jk} \Gamma^\lambda_{\lambda\mu}+
\varphi^jF^{\mu\nu}_{jk,\nu} \Gamma^\lambda_{\lambda\mu}+
\varphi^jF^{\mu\nu}_{jk} \Gamma^\lambda_{\lambda\mu,\nu}+\nonumber\\
& &\left.\qquad \varphi^jF^{\mu\nu}_{jk} \Gamma^\lambda_{\lambda\mu}\Gamma^\rho_{\rho\nu}\right]
\psi^kJd^4\sigma.
\label{byparts2}
\end{eqnarray}
By substituting  (\ref{byparts1})-(\ref{byparts2}) into (\ref{symmetry1}) 
and assuming that the field variations $\varphi^j$ and $\psi^k$ 
are arbitrary, we obtain that the formal symmetry condition 
is satisfied if and only if
\begin{eqnarray}
F^{\mu\nu}_{kj}&=&F^{\mu\nu}_{jk},\label{zerocondition}\\
B^\nu_{kj}&=&-B^\nu_{jk}+2\left(F^{\mu\nu}_{jk,\mu}+
F^{\mu\nu}_{jk}\Gamma^\lambda_{\lambda\mu}\right),\label{firstcondition}\\
H_{kj}&=&H_{jk}+F^{\mu\nu}_{jk,\mu\nu}+2 F^{\mu\nu}_{jk,\mu}\Gamma^\lambda_{\lambda\nu}+
F^{\mu\nu}_{jk}\Gamma^\lambda_{\lambda\mu,\nu}+
F^{\mu\nu}_{jk}\Gamma^\lambda_{\lambda\mu}\Gamma^\rho_{\rho\nu}-
B^\nu_{jk,\nu}-B^\nu_{jk}\Gamma^\lambda_{\lambda\nu}.\label{secondcondition}
\end{eqnarray}
This is a system of nonlinear PDEs 
that, in principle, allows us to identify the functional relation between 
the conjugate flow $\x^\mu$ and the  field $u^j$, 
i.e., the {\em functional manifold} on which the vector field theory is potential. 
We shall call Eqs. (\ref{zerocondition})-(\ref{secondcondition}) the 
{\em determining equations} of the conjugate flow.
It is interesting to note that if 
we remove the functional link between $\x^\mu$ and $u^j$, and 
we consider fixed Cartesian coordinates then 
the conditions (\ref{zerocondition})-(\ref{secondcondition}) 
consistently reduce to those of Tonti \cite{Tonti2}.

\subsection{Scalar field theories}
Let us consider a scalar field theory governed by 
the nonlinear second-order PDE
\begin{equation}
\N_{\x}(u)=f\left(u; u_{,\mu}; u,_{\mu\nu}; \x^\mu_{,\nu}; \x^\mu_{,\nu\lambda} \right)=0.
\label{opeq_s}
\end{equation}
A simple example of (\ref{opeq_s}) is the intrinsic form of the 
diffusion equation (\ref{ffr}). 
The operator $\G_{\x}$  in this case is obtained as 
\begin{equation}
 \G_{\x}\varphi=Q\varphi + Z^\nu \varphi_{,\nu}+R^{\mu\nu}\varphi_{,\mu\nu},\label{GGs}
\end{equation}
where 
\begin{eqnarray}
Q&\Def&\frac{\partial f}{\partial u}+\frac{\partial f}{\partial \x^\mu_{,\nu}}b^\mu_\nu
+\frac{\partial f}{\partial \x^\mu_{,\nu\rho}}c^\mu_{\nu\rho},\\
Z^\nu&\Def& \frac{\partial f}{\partial u_{,\nu}}+
\frac{\partial f}{\partial \x^\mu_{,\nu}}a^\mu+
\left(\frac{\partial f}{\partial \x^\mu_{,\nu\rho}}+
\frac{\partial f}{\partial \x^\mu_{,\rho\nu}}\right)b^\mu_\rho,\label{Bs}\\
R^{\rho\nu}&\Def&\frac{\partial f}{\partial u_{,\nu\rho}}+
\frac{\partial f}{\partial \x^\mu_{,\rho\nu}}a^\mu, \label{Fs}
\end{eqnarray}
and 
\begin{eqnarray}
a^\mu\Def\frac{\partial \x^\mu}{\partial u},\qquad 
b^\mu_\nu\Def a^\mu_{,\nu}+\frac{\partial a^\mu}{\partial u}u_{,\nu},\qquad
c^\mu_{\nu\rho}\Def b^\mu_{\nu,\rho}+\frac{\partial b^\mu_{\nu}}{\partial u}u_{,\rho}.
\end{eqnarray}
By using the local inner product (\ref{local_bilinear_form}) 
we write explicitly the symmetry condition (\ref{er}) as
\begin{eqnarray}
 \int_{\Sigma} \left(\psi Q\varphi +\psi Z^\mu\varphi_{,\mu} + 
 \psi R^{\mu\nu}\varphi_{,\mu\nu}\right) Jd^4\sigma=
 \int_{\Sigma} \left(\varphi Q\psi +\varphi Z^\mu\psi_{,\mu} + 
 \varphi R^{\mu\nu}\psi_{,\mu\nu}\right) Jd^4\sigma.\label{symmetry2}
\end{eqnarray}
At this point, we integrate by parts the terms at the right hand 
side and neglect all the boundary contributions. 
This yields, for arbitrary field variations $\varphi$ and 
$\psi$
\begin{eqnarray}
Z^\mu&=&R^{\mu\nu}_{,\nu}+R^{\mu\nu}\Gamma^\lambda_{\lambda\nu},\label{firstcondition_s}\\
Z^\mu_{,\mu}&=&R^{\mu\nu}_{,\mu\nu}+2F^{\mu\nu}_{,\nu}\Gamma^\lambda_{\lambda\mu}+
R^{\mu\nu}\Gamma^\lambda_{\lambda\mu}\Gamma^\rho_{\rho\nu}+
R^{\mu\nu}\Gamma^\lambda_{\lambda\mu,\nu}-Z^\nu\Gamma^\lambda_{\lambda\nu}.\label{secondcondition_s}
\end{eqnarray}
A differentiation of Eq. (\ref{firstcondition_s}) with respect 
to $\sigma^\mu$ and subsequent substitution into
Eq. (\ref{secondcondition_s}) gives the single relation
\begin{equation}
\left(R^{\mu\nu}_{,\nu}+R^{\mu\nu}\Gamma^\lambda_{\lambda\mu}-Z^\nu\right)\Gamma^\mu_{\mu\nu}=0,
\end{equation}
which is equivalent to the following system of \emph{determining equations} for 
the conjugate flow
\begin{equation}
R^{\mu\nu}_{,\nu}+R^{\mu\nu}\Gamma^\lambda_{\lambda\nu}=Z^\nu. \label{final_condition}
\end{equation}
This system defines the functional 
relation between the field $u$ and the flow $\x^\mu$ 
for which the scalar field theory is potential.
Note that if we remove the functional link 
between $\x^\mu$ and $u$, then the conditions (\ref{final_condition}) consistently 
reduce to those of Tonti \cite{Tonti1,Tonti2}. In order to see this, we simply 
set $a^\mu$, $b^\mu_\nu$ equal to zero in Eq. (\ref{Bs}) and Eq. (\ref{Fs}) 
and then substitute them into Eq. (\ref{final_condition}). The result 
in fixed Cartesian coordinates ($\Gamma^\beta_{\beta\rho}=0$) is
\begin{equation}
\frac{\partial }{\partial x^\mu}\left(\frac{\partial f}{\partial u_{,\mu\nu}}\right)
-\frac{\partial f}{\partial u_{,\nu}}=0.
\end{equation}
This is the classical condition arising from the symmetry 
requirement of a second-order nonlinear scalar field equation \cite{Filippov,Finlayson}.

\begin{table}
\begin{tabular}{c|l}
\begin{minipage}{7cm}
Field Equation  
\end{minipage}
&\hspace{0cm}
Determining Equations for the Conjugate Flow
\\
\hline\\
$f\left(u; u_{,\mu}; u,_{\mu\nu}; \x^\mu_{,\nu}; \x^\mu_{,\nu\lambda} \right)=0$ & 
$R^{\mu\nu}_{,\nu}+R^{\mu\nu}\Gamma^\lambda_{\lambda\nu}=Z^\nu$ \\\\
 \hline
$f_k\left(u^j; u^j_{,\mu}; u^j_{,\mu\nu}; \x^\mu_{,\nu}; \x^\mu_{,\nu\lambda} \right)=0$ & 
\begin{minipage}{6cm}
\begin{eqnarray}
F^{\mu\nu}_{kj}&=&F^{\mu\nu}_{jk}\nonumber\\
B^\nu_{kj}&=&-B^\nu_{jk}+2\left(F^{\mu\nu}_{jk,\mu}+
F^{\mu\nu}_{jk}\Gamma^\lambda_{\lambda\mu}\right),\nonumber\\
H_{kj}&=&H_{jk}+F^{\mu\nu}_{jk,\mu\nu}+2 F^{\mu\nu}_{jk,\mu}\Gamma^\lambda_{\lambda\nu}+
F^{\mu\nu}_{jk}\Gamma^\lambda_{\lambda\mu,\nu}+\nonumber\\
& & F^{\mu\nu}_{jk}\Gamma^\lambda_{\lambda\mu}\Gamma^\rho_{\rho\nu}-
B^\nu_{jk,\nu}-B^\nu_{jk}\Gamma^\lambda_{\lambda\nu}.\nonumber
\end{eqnarray}
\end{minipage}\\\\\hline
\end{tabular}
\caption{Determining equations for the conjugate flow that symmetrizes  
second-order scalar and vector field theories. Once the conjugate flow is available, 
the Lagrangian density and the action functional can be determined 
by calculating the integral (\ref{lagrangian}) along an arbitrary trajectory 
of fields.}
\label{tab:1}
\end{table}

\section{Summary}
\label{sec:summary}
We developed a new general approach to construct an action 
functional for a non-potential field theory. The key idea 
relies on representing the governing equations of the theory 
relative to a functional flow of curvilinear coordinates 
(the conjugate flow) which is assumed to be a dependent on 
the solution field. We have shown that such flow 
can be selected in order to symmetrize the G\^ateaux 
derivative of the field equations with respect to suitable local 
bilinear forms. This is equivalent to 
requiring that the field equations of the theory can 
be derived from a principle of stationary 
action on a Lie group manifold. By using a general operator 
framework, we obtained the determining equations of such
manifold for second-order scalar and vector field theories 
and shown that they are consistent with classical results in 
fixed coordinates. Once the symmetrizing conjugate 
flow is available, the action functional of the theory can 
be constructed explicitly by using path integration. 
In particular, the duality principle between the conjugate 
flow and the solution field allows us to perform integrations either 
in terms of flows or in terms of fields, yielding two different 
types of actions.

The proposed new methodology can be applied to scalar, vector and tensor 
field theories. In particular, it can be applied to the Navier-Stokes 
equations, for which a great research effort has focused in obtaining 
a physically meaningful principle of stationary 
action \cite{Cipriano,Kerswell,Mobbs,Finlayson1}. 
Recent results of Gomes \cite{Gomes,Gomes1} and Eyink \cite{Eyink} 
indeed have shown that an action principle can be constructed 
on {\em random diffeomorphisms} \cite{Funaki,Yasue,Rapoport}. 
These random flows are usually defined in terms of stochastic 
perturbations of Lagrangian base flows. 
By using the methods of the present paper, the variational principle 
for the Navier-Stokes equations may be constructed on a 
deterministic functional diffeomorphism, i.e., on a 
conjugate flow satisfying the set of determining 
equations summarized in table 2. 
In fact, the representation of 
the Navier-Stokes equations in terms of a symmetrizing 
conjugate flow yields a potential field theory which 
admits a Lagrangian density and an action functional. 
These quantities can be explicitly determined by using 
the methods of section \ref{sec:action}. In particular, 
once the conjugate flow is available, the Lagrangian density 
can be determined by computing the integral (\ref{lagrangian}) 
along a trajectory of velocity fields. 
The identification of groups of transformations leaving 
the conjugate flow action functional invariant could lead 
to the discovery of new conservation laws in fluid dynamics 
and other disciplines.

\appendix
\section{Representation of field equations in conjugate flow intrinsic coordinates}
\label{app:B}
In this appendix we recall some fundamental identities of differential 
geometry \cite{Lovelock,Aris,Truesdell1,Landau} that allow us 
to write the dynamic equations of a physical system in terms of conjugate flow 
intrinsic coordinates. To this end, let us first consider 
the Jacobian of the conjugate flow transformation (\ref{tr_conj})
\begin{equation}
 J^\mu_\nu\Def\frac{\partial \x^\mu }{\partial \sigma^\nu} \quad\quad\mu,\nu=0,..,n,
 \label{Jako}
\end{equation}
where $n$ is the number of spatial dimensions and $0$ denotes the 
temporal component. 
It is easy to verify that the transpose of the algebraic 
complement of $J^\mu_\nu$ has tensorial expression 
(repeated indices are summed)
\begin{equation}
C^\lambda_\rho\Def\frac{1}{n!}\epsilon^{\lambda\nu\alpha\cdots}
\epsilon_{\rho\mu\lambda\cdots}
\frac{\partial \widehat{x}^\mu}{\partial \sigma^\nu}
\frac{\partial \widehat{x}^\lambda}{\partial \sigma^\alpha}\cdots,\label{C}
\end{equation}
where $\epsilon^{\lambda\nu\alpha\cdots}$ 
and $\epsilon_{\rho\mu\lambda\cdots}$ are multi-dimensional 
permutation symbols, i.e. Levi-Civita tensorial densities. For example, if 
we consider one spatial and one temporal dimension 
then we obtain the simple expression (all indices are from $0$ to $1$)
\begin{equation}
C^\lambda_\rho=\epsilon^{\lambda\nu}\epsilon_{\rho\mu}
\frac{\partial \widehat{x}^\mu}{\partial \sigma^\nu}.\label{C2}
\end{equation}
Similarly, in $1+3$ dimensions, i.e. one temporal and three spatial 
dimensions (all indices are from $0$ to $3$)
\begin{equation}
C^\lambda_\rho=\frac{1}{6}\epsilon^{\lambda\nu\alpha\beta}
\epsilon_{\rho\mu\lambda\delta}
\frac{\partial \widehat{x}^\mu}{\partial \sigma^\nu}
\frac{\partial \widehat{x}^\lambda}{\partial \sigma^\alpha}
\frac{\partial \widehat{x}^\delta}{\partial \sigma^\beta}.\label{C4}
\end{equation}
By using Eqs. (\ref{Jako}) and (\ref{C}) we obtain the 
Jacobian determinant
\begin{equation}
J\Def\det\left(J^\mu_\nu\right)=\frac{1}{n+1}J^\mu_\nu C^\nu_\mu.\label{J}
\end{equation}
This allows us to write the the inverse of the Jacobian 
matrix (\ref{Jako}) as  
\begin{equation}
 A^\lambda_\rho\Def\frac{C^\lambda_\rho}{J}.
 \label{A}
\end{equation}
At this point, let us denote by
\begin{equation}
 \sigma^\nu=\s^\nu\left(x^\mu;u^j\right),\quad\quad\mu,\nu=0,..,n
 \label{coord trans1}
\end{equation} 
the inverse transformation of (\ref{tr_conj}). Such inverse 
transformation exists and it is differentiable by definition of conjugate flow.
From the well known identity
\begin{equation}
\frac{\partial \widehat{\sigma}^\nu}{\partial x^\mu}
\frac{\partial \widehat{x}^\mu}{\partial \sigma^\lambda}=\delta^{\nu}_\lambda
\end{equation}
we obtain the following \emph{fundamental expression} of the partial 
derivatives $\partial \widehat{\sigma}^\nu/ \partial x^\mu$ as 
a function of $\sigma^\lambda$
\begin{equation}
 \frac{\partial \widehat{\sigma}^\nu}{\partial x^\mu}=
 A^\nu_\mu\left(\sigma^\lambda;u^j\right),\label{fondrel}
\end{equation}
where $A^\nu_\mu$ is defined in Eq. (\ref{A}).
It is useful to write down $A^\nu_\mu$ explicitly 
in the two-dimensional case
\begin{equation}
\left[
\begin{array}{cc}
A^0_0 & A^0_1\\
A^1_0 &A^1_1
\end{array}
\right]=\frac{1}{J}\left[
\begin{array}{cc}
 \partial \x^1/\partial\sigma^1  & -\partial \x^0/\partial \sigma^1\\
-\partial \x^1/\partial\sigma^0  &  \partial \x^0/\partial \sigma^0
\end{array}\right],\label{the A}
\end{equation}
where 
\begin{equation}
J=\frac{\partial \x^1}{\partial \sigma^1}\frac{\partial \x^0}{\partial \sigma^0}-
\frac{\partial \x^0}{\partial \sigma^1}\frac{\partial \x^1}{\partial \sigma^0}.
\end{equation}
If time is not transformed, i.e. if  $x^0=\sigma^0=t$, then Eq. (\ref{the A}) 
reduces to
\begin{equation}
\left[
\begin{array}{cc}
A^0_0 & A^0_1\\
A^1_0 &A^1_1
\end{array}
\right]=\left[
\begin{array}{cc}
1 & 0\\
-(\partial \x/\partial t)/(\partial \x/\partial \sigma) & 
1/(\partial \x/\partial \sigma)
\end{array}
\right],\label{the A1}
\end{equation}
where we have denoted by $\sigma\equiv\sigma^1$ and $x\equiv x^1$. 

\subsection*{Partial differentiation in intrinsic coordinates}
Let us consider a vector field  $U^j\left(x^\mu\right)$ 
in fixed Cartesian coordinates $x^\mu$. Such field 
can be equivalently expressed relative to conjugate flow 
intrinsic coordinates by using the transformation (\ref{tr_conj}). 
This yields the following equivalent representations
 \begin{equation}
 U^j\left(x^\mu\right)=U^j\left(\x^\mu\left(\sigma^\nu\right)\right)=
 u^j\left(\sigma^\nu\right)=u^j\left(\s^\nu\left(x^\mu\right)\right).\label{transa}
\end{equation}
The transformation law for partial derivatives of $U^j$ 
is easily obtained by differentiating Eq. (\ref{transa}) 
\begin{equation}
\frac{\partial U^j}{\partial x^\mu}=\frac{\partial u^j}{\partial \sigma^\nu}
\frac{\partial \widehat{\sigma}^\nu}{\partial x^\mu}
=\frac{\partial u^j}{\partial \sigma^\nu}A^\nu_\mu,\label{covariant}
\end{equation} 
where the quantities $\partial \s^\nu/\partial x^\mu$ are expressed 
in coordinates $\sigma^\nu$ through Eq. (\ref{fondrel}). 
Let us now evaluate the second derivative with respect to $x^\nu$ and 
express the result in conjugate flow intrinsic coordinates. To this end 
let us perform an additional differentiation of (\ref{covariant}) with 
respect to $x^\nu$. This yields
\begin{eqnarray}
\frac{\partial^2 U^j}{\partial x^\mu\partial x^\nu}&=&\frac{\partial^2u^j}{\partial \sigma^\lambda\partial \sigma^\rho}
A^\rho_\nu A^\lambda_\mu+\frac{\partial u^j}{\partial \sigma^\lambda}\frac{\partial A^\lambda_\mu}{\partial \sigma^\rho}A^\rho_\nu .\label{second}
\end{eqnarray}
By using the expressions of $J$ and  $C^\nu_\mu$ obtained in (\ref{J}) 
and (\ref{C}) it is possible to manipulate Eq. (\ref{second}) further. 
However, it is more convenient to obtain first $A^\nu_\mu$ explicitly as a 
function of $\sigma^\mu$ and then perform the differentiation appearing in 
(\ref{second}).

\subsection*{Perturbations of the metric tensor, affine connection and Jacobian determinant}
When the conjugate flow (\ref{tr_conj}) undergoes an infinitesimal disturbance 
of type (\ref{trajectory_perturbation}) then all the quantities 
related to its intrinsic geometry are subject to small 
variations. For instance, the metric tensor
\begin{equation}
g_{\mu\nu}\Def\frac{\partial \x^\beta}{\partial \sigma^\mu}
\frac{\partial\x^\beta}{\partial \sigma^\nu}.\label{themetric}
\end{equation}
becomes, to the first-order in $\epsilon$, $g_{\mu\nu}+\epsilon h_{\mu\nu}$ where
\begin{eqnarray}
 h_{\mu\nu}&\Def&\frac{\partial \widehat{\phi}^\beta}{\partial \sigma^\mu}
 \frac{\partial\x^\beta}{\partial \sigma^\nu}+
 \frac{\partial \x^\beta}{\partial \sigma^\mu}
 \frac{\partial\widehat{\phi}^\beta}{\partial \sigma^\nu}.
\end{eqnarray}
The corresponding perturbation in the affine connection
(Christoffel symbols of the second kind) 
\begin{eqnarray}
\Gamma^\alpha_{\mu\nu}&\Def&\frac{\partial \widehat{\sigma}^\alpha}{\partial x^\rho}
\frac{\partial^2 \x^\rho}{\partial \sigma^\mu\partial \sigma^\nu} \label{affine}
\end{eqnarray}
can be obtained by using the well-known representation in terms of 
metric tensor, i.e., 
\begin{equation}
\Gamma^\alpha_{\mu\nu}=\frac{1}{2}g^{\alpha\rho}\left(\frac{\partial g_{\rho\mu}}{\partial \sigma^\nu}+
\frac{\partial g_{\rho\nu}}{\partial \sigma^\mu}-
\frac{\partial g_{\mu\nu}}{\partial \sigma^\rho}\right).
\end{equation}
This yields\cite{Weinberg}
\begin{eqnarray}
\delta\Gamma^\alpha_{\mu\nu}&=&-\epsilon g^{\alpha\rho}h_{\rho\beta}\Gamma^{\beta}_{\mu\nu}+
\frac{\epsilon}{2}g^{\alpha\rho}\left(\frac{\partial h_{\rho\mu}}{\partial \sigma^\nu}+
\frac{\partial h_{\rho\nu}}{\partial \sigma^\mu}-
\frac{\partial h_{\mu\nu}}{\partial \sigma^\rho}\right)\nonumber\\
&=&\frac{\epsilon}{2}g^{\alpha\rho}\left(h_{\rho\mu;\nu}+h_{\rho\nu;\mu}-h_{\mu\nu;\rho}\right)
\label{perturbation_affine_c}
\end{eqnarray}
the covariant derivatives ``$;$'' being of course constructed by
using the unperturbed affine connection $\Gamma^\alpha_{\mu\nu}$.
These results allows us to compute the conjugate flow perturbation of 
other fundamental geometric quantities such as the Riemann-Christoffel 
curvature tensor and the Jacobian determinant (\ref{J}) of the conjugate 
flow transformation. To this end, we substitute Eq. (\ref{trajectory_perturbation}) 
into Eq. (\ref{J}) and we keep only the linear terms in $\epsilon$ to obtain 
\begin{eqnarray}
J+\epsilon C^\nu_\mu\frac{\partial \widehat{\phi}^\mu}{\partial \sigma^\nu}
=J\left(1+ \epsilon\frac{\partial \widehat{\phi}^\mu}{\partial x^\mu}\right).\label{perturbedJ}
\end{eqnarray}
The quantity $\partial \widehat{\phi}^\mu/\partial x^\mu$ is the divergence 
of the flow perturbation (remember that $\widehat{\phi}^\mu$ are 
Cartesian components). 
Another useful formula involving the Jacobian 
determinant is \cite{Lovelock,Weinberg}
\begin{equation}
\frac{\partial J}{\partial \sigma^\mu}=J\Gamma^\nu_{\nu\mu}.
\label{derJac}
\end{equation}


\end{document}